\documentclass[a4paper,12pt]{article}
\usepackage{a4wide}
\usepackage[utf8]{inputenc}
\usepackage{graphicx}
\usepackage{amsmath}
\usepackage{amssymb}
\usepackage{hyperref}
\allowdisplaybreaks[1]

\begin{document}

\begin{titlepage}
\begin{flushright}
LU TP 17-29\\
October 2017 \\
\end{flushright}
\vfill
\begin{center}

{\large\bf The Pion Mass and Decay Constant at Three Loops\\[2mm] in Two-Flavour Chiral Perturbation Theory}
\\[3cm]
{\bf Johan Bijnens and Nils Hermansson Truedsson}
\\[5mm]
{Department of Astronomy and Theoretical Physics, \\Lund University, S\"{o}lvegatan 14A, SE 223-62 Lund, Sweden }
\vfill

\large{\textbf{Abstract}}
\end{center}

A calculation of the pion mass and decay constant at NNNLO in two-flavour
chiral perturbation theory is presented. The results are cross-checked by
using both the exponential and square root parameterizations of the Goldstone
matrix field, as well as by comparing to the known leading log coefficients
of the two quantities. A small numerical study of the quark mass dependence
is performed, and for a physical quark mass there is good agreement with
lower order results.
\end{titlepage}


\section{Introduction}
\label{IntroSec}

Chiral perturbation theory (ChPT) 
\cite{Weinberg:1978kz,Gasser:1983yg}
is a low energy effective field theory of QCD. It is built using the
approximate chiral symmetry
$SU\left( N_{f}\right) _{L}\times SU\left( N_{f}\right) _{R}$ of QCD,
where $N_{f}$ is the number of quark flavours, which is spontaneously broken
to $SU\left( N_{f}\right)_{V}$ by a non-vanishing quark
condensate $\langle \bar{q}q\rangle \neq 0$. The $N_{f}^2-1$ broken generators
yield as many pseudo-Goldstone bosons. These are identified with the
lightest pseudoscalar mesons living in the coset space
$SU\left( N_{f}\right) _{L}\times SU\left( N_{f}\right) _{R}/SU(N_{f})_{V}$. 
For $N_F=2$ or  $SU(2)$ only pions appear, whereas for $N_f=3$ or $SU(3)$
there are the pions, kaons and the eta.

The masses and decay constants of these composite particles can be calculated
within ChPT to a given order in the chiral expansion, i.e., to order
$\left( p^2\right) ^n$ where the integer $n\geq 1$. These were known at
next-to-next-to-leading order (NNLO)
for the pions in both $SU(2)$~\cite{Burgi:1996qi,Bijnens:1995yn,Bijnens:1997vq}
and $SU(3)$~\cite{Amoros:1999dp}.
In this paper we extend the two-flavour or $SU(2)$ case to the next order,
$p^8$ or NNNLO.
All relevant three-loop integrals are known
\cite{Laporta:1996mq,Melnikov:2000zc}.
As a consistency check, we also calculate the mass at three-loop
order in $O(N)$ $\phi ^{4}$ theory.
The general method to NNLO is described in detail in \cite{Bijnens:1997vq}. We
extend it to one order higher in the expansion.

The motivation behind this work is twofold. The expressions themselves are of
intrinsic interest but in so-called hard-pion ChPT it was argued that mass
logarithms could be calculated also for pions with hard momenta.
This was checked at two-loop order in ChPT \cite{Bijnens:2010jg}.
At three-loop order it was found that the chiral mass logarithm does not agree
with the prediction of \cite{Bijnens:2010jg} in \cite{Colangelo:2012ew}.
This work is a first step towards checking the results
of \cite{Colangelo:2012ew} and possibly being able to correct and extend the
arguments of \cite{Bijnens:2010jg} towards a full proof.

\section{Mass in the $O(N)$  $\phi^4$ model}
\label{sec:phi4}

The Lagrangian for the $O(N)$  $\phi^4$ model is given by
\begin{align}
\label{Lagphi4}
\mathcal{L} =\,& 
\left(1+l_1\lambda+c_1\lambda^2+d_1\lambda^3\right)\frac{1}{2}
  \partial_\mu\phi^T\partial^\mu\phi
-\left(1+l_2\lambda+c_2\lambda^2+d_2\lambda^3\right)\frac{1}{2}M^2
  \phi^T\phi
\nonumber\\ &
-\left(1+l_3\lambda+c_3\lambda^2+d_3\lambda^3\right)\frac{\lambda}{4}
  \left(\phi^T\phi\right)^2-\phi^T f\,.
\end{align}
$\phi$ is a vector of
$N$ real fields $\phi_a$ and $f$ is the external current. We have indicated
here the higher order terms with $c_i,l_i$ and $d_i$ as well and have used the
equations of motion (or field redefinitions) to discard the higher order terms
in the coupling to the external current $f$. Up to two-loop order the
counterterms are given by\footnote{These are slightly different from
\cite{Bijnens:1997vq} since we have  chosen not to put $l_1^r=0$.}
\begin{align}
\label{subtractionphi4}
l_1 =\,& (c\mu)^{-2\varepsilon}\left(l_1^r\right) 
\nonumber\\
l_2 =\,& (c\mu)^{-2\varepsilon}\left(l_2^r+\frac{N+2}{16\pi^2\varepsilon}\right)
\nonumber\\
l_3 =\,& (c\mu)^{-2\varepsilon}\left(l_3^r+\frac{N+8}{16\pi^2\varepsilon}\right)
\nonumber\\
c_1 =\,& (c\mu)^{-4\varepsilon}\left(c_1^r+\frac{1}{(16\pi^2)^2\varepsilon}\left(-\frac{N+2}{2}\right)\right)
\nonumber\\
c_2 =\,& (c\mu)^{-4\varepsilon}\left(c_2^r+\frac{(N+5)(N+2)}{(16\pi^2\varepsilon)^2}
-\frac{3(N+2)}{(16\pi^2)^2\varepsilon}
+\frac{2(N+2)}{16\pi^2\varepsilon}\left(-2l_1^r+l_2^r+l_3^r\right)
\right)
\nonumber\\
c_3 =\,& (c\mu)^{-4\varepsilon}\left(c_3^r+\frac{(N+8)^2}{(16\pi^2\varepsilon)^2}
-\frac{2(5N+22)}{(16\pi^2)^2\varepsilon}
+\frac{2(N+8)}{16\pi^2\varepsilon}\left(-l_1^r+l_3^r\right)
\right)
\end{align}
We use dimensional regularization with $d=4-2\varepsilon$ and modified minimal
subtraction ($\overline{MS}$). The choice of $c$ determines which version
of $\overline{MS}$ is used. In this manuscript we use the usual ChPT \cite{Gasser:1983yg} version with
\begin{align}
c = -\frac{1}{2}\left[\log(4\pi)+\Gamma^\prime(1)+1\right]\,.
\end{align}

The mass is defined as the pole of the two point function, see e.g. the
discussion in \cite{Amoros:1999dp}, as
\begin{align}
\label{MassProp}
\frac{i}{p^{2}-M^{2}-\Sigma \left(p^2 \right) }
\end{align}
where $\Sigma $ is the sum of one-particle irreducible diagrams.
The relevant diagrams are shown in Fig.~\ref{DiagramFig} when neglecting those
involving vertices with more than four legs. $p^2$ corresponds to terms
with 1, $p^4$ with $l_i$, $p^6$ with $c_i$ and $p^8$ with $d_i$ in
(\ref{Lagphi4}).

The physical mass $M_\phi$ is given by the solution of
\begin{align}
\label{solvemass}
M_\phi^2-M^2-\Sigma(M_\phi^2)=0\,.
\end{align}
We write the mass as
\begin{align}
M_\phi^2 = M^2\left(1+M_4^2+M_6^2+M_8^2\right)
\end{align}
and the self-energy as
\begin{align}
\Sigma(p^2) = \Sigma_4(p^2)+\Sigma_6(p^2)+\Sigma_8(p^2)\,.
\end{align}
Where we used the subscript 4,6,8 for NLO, NNLO and NNNLO respectively.
With this expansion we can solve for the mass perturbatively.
We evaluate diagrams at $p^2=M_\phi^2$ as a perturbative expansion away
from $M^2$. This is why derivatives of the self-energy show up.
Taking into account that here $\partial^2 \Sigma_4/(\partial p^2)^2=0$ we get
\begin{align}
M^2 M_4 =\,& \Sigma_4(M^2)
\nonumber\\
M^2 M_6 =\,& \Sigma_6(M^2)+M^2 M_4^2\frac{\partial \Sigma_4}{\partial p^2}
\nonumber\\
M^2 M_8 =\,& \Sigma_8(M^2)+M^2 M_6^2\frac{\partial \Sigma_4}{\partial p^2}
  + M^2 M_4^2\frac{\partial\Sigma_6}{\partial p^2}(M^2)\,.
\end{align} 
All nonlocal divergences cancel as they should and we get a finite result
by setting
\begin{align}
\label{d2md1inf}
d_2-d_1 =\,&(c\mu)^{-6\varepsilon}\Bigg\{ d_2^r-d_1^r
        -\frac{\pi_{16}^3}{\varepsilon^3}\left(- 60 - 52 N - 13 N^2 - N^3\right)
\nonumber\\&
        -\frac{\pi_{16}^3}{\varepsilon^2}\left(\frac{284}{3} + \frac{206}{3}N
                                              +\frac{ 32}{3}N^2\right)
        -\frac{\pi_{16}^3}{\varepsilon}\left( - 74 - 47 N - 5 N^2\right)
\nonumber\\&
        -\frac{\pi_{16}^2}{\varepsilon^2}\left(- 20l_3^r - 10 l_2^r - 14 N l_3^r
        - 7Nl_2^r - 2N^2l_3^r - N^2l_2^r + 40l_1^r + 28Nl_1^r + 4N^2l_1^r\right)
\nonumber\\&
        -\frac{\pi_{16}^2}{\varepsilon}\left(10l_3^r + 6l_2^r + 5Nl_3^r
             + 3Nl_2^r - 21l_1^r - \frac{21}{2}Nl_1^r\right)
\nonumber\\&
        -\frac{\pi_{16}}{\varepsilon}( - 2c_3^r - 2c_2^r + 4c_1^r
         - 2l_2^rl_3^r - Nc_3^r - Nc_2^r + 2Nc_1^r - Nl_2^rl_3^r +4l_1^rl_3^r
              + 4l_1^rl_2^r
\nonumber\\&
        \hskip1.5cm - 6{l_1^r}^2 + 2Nl_1^rl_3^r + 2Nl_1^rl_2^r - 3N{l_1^r}^2
          )
\Bigg\}\,.
\end{align}
Here we introduced the shorthand $\pi_{16}=1/(16\pi^2)$.
We express the result for the mass in terms of the logarithm
\begin{align}
L_M = \log\frac{M^2}{\mu^2}\,.
\end{align}
The full result for the mass at three-loop order is
\begin{align}
M_4^2/\lambda =\,& (N+2)\pi_{16}L_M+l_2^r-l_1^r
\nonumber\\
M_6^2/\lambda^2 =\,&
        c_2^r - c_1^r - l_1^rl_2^r + {l_1^r}^2 
       + \pi_{16} ( 2l_2^r - 2l_1^r + Nl_2^r - Nl_1^r )
\nonumber\\&
       + \pi_{16} L_M  ( 2l_3^r + 2l_2^r - 6l_1^r + Nl_3^r + Nl_2^r
          - 3Nl_1^r )
       + \pi_{16}^2 ( 3/2 + 3/4N )
\nonumber\\&
       + \pi_{16}^2L_M  (  - 6 - N + N^2 )
       + \pi_{16}^2L_M^2  ( 10 + 7N + N^2 )
\nonumber\\
M_8^2/\lambda^3 =\,&
       d_2^r-d_1^r - l_2^rc_1^r - l_1^rc_2^r + 2l_1^rc_1^r + {l_1^r}^2l_2^r
          - {l_1^r}^3 
\nonumber\\&
       + \pi_{16} (N+2) ( c_2^r - c_1^r + l_2^rl_3^r + (1/2){l_2^r}^2 - l_1^rl_3^r
          - 4l_1^rl_2^r + (7/2){l_1^r}^2  )
\nonumber\\&
       + \pi_{16}L_M(N+2)  ( c_3^r + c_2^r - 3c_1^r + l_2^rl_3^r - 3
         l_1^rl_3^r - 3l_1^rl_2^r + 6{l_1^r}^2 )
\nonumber\\&
       + \pi_{16}^2  ( 3l_3^r - (9/2)l_2^r - (3/2)l_1^r + (3/2)Nl_3^r - (1/4)
         Nl_2^r - (11/4)Nl_1^r + N^2l_2^r - N^2l_1^r )
\nonumber\\&
       + \pi_{16}^2L_M  (  - 12l_3^r + 14l_2^r + 10l_1^r - 2Nl_3^r
          + 13Nl_2^r - 9Nl_1^r + 2N^2l_3^r + 3N^2l_2^r - 7N^2l_1^r )
\nonumber\\&
       + \pi_{16}^2L_M^2  ( 20l_3^r + 10l_2^r - 50l_1^r + 14Nl_3^r
          + 7Nl_2^r - 35Nl_1^r + 2N^2l_3^r + N^2l_2^r - 5N^2l_1^r )
\nonumber\\&
       + \pi_{16}^3  ( 124/3 + 64\zeta_3 + (239/6)N + 40N\zeta_3 + (115/12)
         N^2 + 4N^2\zeta_3 )
\nonumber\\&
       + \pi_{16}^3L_M  ( 217 + 131N + (53/4)N^2 + N^3 )
\nonumber\\&
       + \pi_{16}^3L_M^2  (  - 128 - 64N + 5N^2 + (5/2)N^3 )
       + \pi_{16}^3L_M^3  ( 60 + 52N + 13N^2 + N^3 )
\end{align}

This result can be checked in a number of ways. The nonlocal divergences
cancelled as they should. The terms leading in $N$ can be derived using a
gap equation similar to what was done for the nonlinear sigma model in
\cite{Bijnens:2009zi,Bijnens:2010xg}. The
renormalization group equations are known to five loop order
\cite{Kleinert:1991rg}, these can be used to check the $1/\varepsilon$ terms in
(\ref{d2md1inf}). All checks are satisfied.

\section{Chiral perturbation theory}

The effective Lagrangian in ChPT is expanded in powers of $p^2$ as
\begin{eqnarray}
\mathcal{L} = \mathcal{L}_{0} +\mathcal{L}_{4} + \mathcal{L}_{6}+\mathcal{L}_{8}+\ldots
\end{eqnarray}
The relevant degrees of freedom are the Goldstone Bosons from the
spontaneous breakdown of $SU(N_f)_L\times SU(N_f)$ to $SU(N_f)_V$. These
can be described by a special unitary $N_f\times N_f$ matrix $u$. For two
flavours the lowest order Lagrangian is
\begin{eqnarray}
\mathcal{L}_{0} = \frac{F^{2}}{4} \langle u_{\mu }u^{\mu }+\chi _{+} \rangle
\end{eqnarray}
with $u_\mu=i\left(u^\dagger(\partial_\mu-i r_\mu)u-u(\partial_\mu-i l_\mu)u^\dagger\right)$, $\chi_+ =u^\dagger\chi u^\dagger+u\chi^\dagger u$, $\chi=B(s+ip)$.
The fields $l_\mu,r_\mu,s$ and $p$ are the usual $N_f\times N_F$ external fields
of ChPT. $F$ and $B$ are the two low-energy-constants (LECs) at leading order
for the two-flavour case. The next-order Lagrangian was classified
in \cite{Gasser:1983yg}. The NNLO Lagrangian can be found
in \cite{Bijnens:1999sh}.  The NNNLO Lagrangian $\mathcal{L}_{8}$ is at present
not known, but there will be one combination of $p^8$ LECs contributing to the
mass and another to the decay constant, we will call these combinations
$r_{M8}$ and $r_{F8}$, respectively.

The NLO and NNLO low-energy-constants (LECs) are conventionally denoted as $l_{i}$ and $c_{i}$, respectively.
The divergent parts needed to one- \cite{Gasser:1983yg} and two-loop order 
\cite{Bijnens:1999hw} are known in general and the equivalent formulas
to (\ref{subtractionphi4}) can be found there. For later convenience we
introduce the lowest order order pion mass
\begin{align}
M^2 = 2 B \hat m
\end{align}
where $2\hat m = m_u+m_d$. In the remainder we will work in the isospin limit
with $m_u=m_d$. 

\section{The calculation and checks}

\begin{figure}
\centerline{
\includegraphics[width=0.7\textwidth]{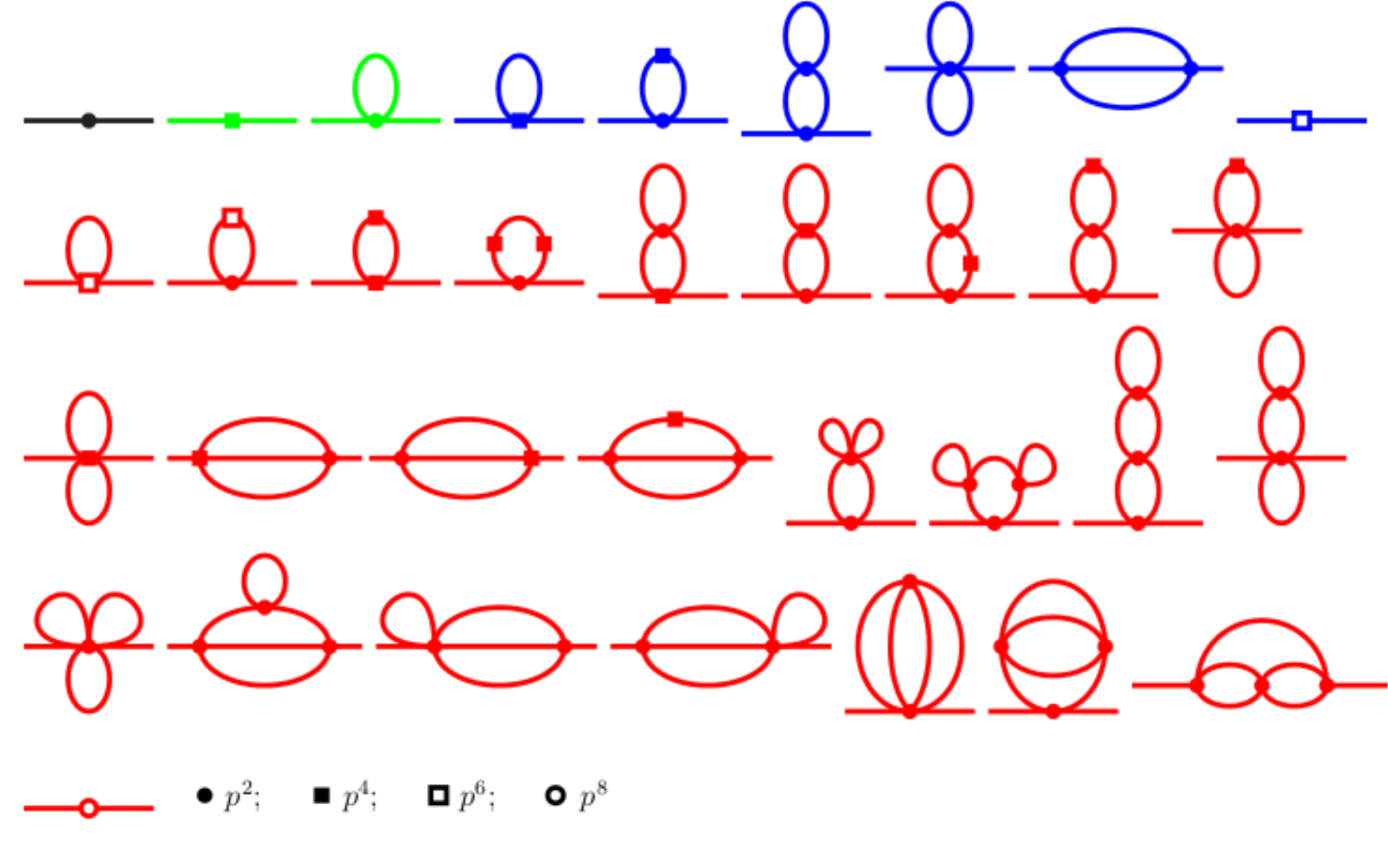}}
\caption{The 34 diagrams contributing to $\Sigma$. The first one is LO, the
next two NLO, the next 6 NNLO, all in the first line. The remaining diagrams
are those needed at NNNLO.}
\label{DiagramFig}
\end{figure}

The diagrams contributing are shown in Fig.~\ref{DiagramFig}.
The Feynman diagrams are programmed in \textsc{FORM} \cite{Vermaseren:2000nd}.
The derivatives w.r.t. $p^2$ needed are obtained by taking
the derivative diagram by diagram at this stage.
Then the expressions are rewritten in integrals. These are reduced
to a set of master integrals. This is done using integration-by-parts
and Lorentz invariance identities through a Laporta algorithm. 
We have used the program \textsc{Reduze}~\cite{Studerus} for this.
The resulting master integrals are all known to the order in $\varepsilon$ required
and we quote them in App.~\ref{IntApp}.

A three-loop calculation needs a large number of checks. We have checked that
the nonlocal divergences cancel, that we reproduce the known two-loop and
leading logarithm results. As a final check we use two different
parametrizations for $u$ in terms of a traceless Hermitian $2\times 2$ matrix
$\Phi$, namely the exponential, $u = \exp \frac{i}{\sqrt{2}F}\Phi$, and
a square root, $u=\sqrt{1-\frac{1}{2F^{2}}\Phi^{2}}+\frac{i}{\sqrt{2}F}\Phi$,
parametrization. The diagrams are quite different in these two parametrizations
but the final result must of course be the same. All checks are satisfied by
our results. Since essentially the same programs were used
for $\phi^4$ the checks discussed in Sect.~\ref{sec:phi4} are another partial
check on our main results.

\section{The pion mass and decay constant}
\label{sec:analytical}

The physical mass is defined as the pole of the two-point
function (\ref{MassProp}) with $\Sigma(p^2)$ the self-energy. The physical
pion mass $M_\pi^2$ is then found as the solution of
\begin{align}
\label{Mpieq}
M_\pi^2-M^2-\Sigma(M_\pi^2)=0\,.
\end{align}
 The decay constant is defined through the relation 
\begin{eqnarray}
\langle 0 | A_{\mu}(0) | \pi (p)\rangle = i\sqrt{2}p_{\mu}F_{\pi}
\end{eqnarray} 
and is calculated using diagrams of the same topology as those in $\Sigma$
(the only difference is that one of the external legs corresponds to the
axial current). For the decay constant one also needs to calculate the
wave function renormalization factor $Z$ defined as the residue of the
propagator in (\ref{MassProp}), i.e.
\begin{eqnarray}
Z=\frac{1}{1-\frac{\partial \Sigma}{\partial p^{2}}}
\end{eqnarray}
at $p^2=M_\pi^2$.

The physical pion mass and decay constant can be written in expanded form
\begin{align}
\label{MassDecExpEqn}
M_{\pi}^2 =\,& M^2\left( 1+M_{4}^{2}+M_{6}^{2}+M_{8}^{2}\right)
\nonumber\\
F_{\pi} =\,& F \left( 1+F_{4}+F_{6}+F_{8}\right) 
\end{align}

\subsection{Mass}

We can solve (\ref{Mpieq}) perturbatively and obtain
\begin{align}
M^2 M_4^2 =\,&\Sigma_4(M^2)
\nonumber\\
M^2 M_6^2 =\,&\Sigma_6(M^2)+M^2 M_4^2\frac{\partial\Sigma_4}{\partial p^2}(M^2)
\nonumber\\
M^2 M_8^2 =\,&\Sigma_8(M^2)+M^2 M_6^2 \frac{\partial\Sigma_4}{\partial p^2}(M^2)
+M^2 M_4^2 \frac{\partial\Sigma_6}{\partial p^2}(M^2)
\end{align}
here we used the fact that $\partial^2 \Sigma_4/(\partial p^2)^2=0$.

In order to obtain a finite result we need the subtraction
\begin{align}
r_{M8} =\,& (c\mu)^{-6\varepsilon}\Big\{ r_{M8}^r
            -\frac{\pi_{16}^3}{\varepsilon^3}(125/72)
            -\frac{\pi_{16}^3}{\varepsilon^2}(71/24)
            -\frac{\pi_{16}^3}{\varepsilon}(28223/12960)
\nonumber\\&
            -\frac{\pi_{16}^2}{\varepsilon^2}(- (7/2)l_4^r + (263/18)l3^r
             - (7/3)l_2^r + (49/3)l_1^r)
\nonumber\\&
            -\frac{\pi_{16}^2}{\varepsilon}(  (302/27)l_3^r + (433/60)l_2^r
                       + (3/20)l_1^r )
\nonumber\\&
            -\frac{\pi_{16}}{\varepsilon}(416c_{18}^r + 208c_{17}^r + 32c_{16}^r
                      - 96c_{14}^r - 8c_{13}^r + 48c_{12}^r + 384c_{11}^r + 
                      192c_{10}^r - 80c_9^r
\nonumber\\&
                      - 160c_8^r - 80c_7^r
                      - 96c_6^r
                      + 8c_5^r + 56c_4^r - 16c_3^r - 32c_2^r
                      + 96c_1^r - 14l_3^r l_4^r + 16 {l_3^r}^2
\nonumber\\&
                      - (16/3)l_2^r l_3^r
                      - (8/3) l_1^r l_3^r )
                \Big\}
\end{align}
$r_{M8}$ is the combination of $p^8$ LECs that contributes to the mass.

This is a single scale problem and only logarithms of the mass scale show up,
the expression is thus fairly compact.
We use the abbreviations
\begin{align}
x =\,& \frac{M^2}{16\pi^2 F^2} & L_M =\,& \log\frac{M^2}{\mu^2}
& l^q_i=\,& 16\pi^2 l^r_i & c^q_i = (16\pi^2)^2 c_i^r
& r^q_{M8} = (16\pi^2)^3 r^r_{M8}\,.
\end{align}
The results can be written in the form
\begin{align}
M_4^2 =\,& x\left(a_{10}^M+a_{11}^M L_M\right)
\nonumber\\
M_6^2 =\,& x^2\left(a_{20}^M+a_{21}^M L_M+a_{22}^M L_M^2\right)
\nonumber\\
M_8^2 =\,& x^3\left(a_{30}^M+a_{31}^M L_M+a_{32}^M L_M^2+a_{33}^M L_M^3\right)
\end{align}

The coefficients are
\begin{align}
a_{10}^M =\,& 2 l_3^q
\nonumber\\
a_{11}^M =\,& 1/2
\nonumber\\
a_{20}^M =\,& 64 c_{18}^q + 32 c_{17}^q + 96 c_{11}^q + 48 c_{10}^q - 16c_9^q
          - 32 c_8^q - 16 c_7^q - 32 c_6^q + l_3^q + 2 l_2^q + l_1^q + 
         (163/96)
\nonumber\\
a_{21}^M =\,& - 3 l_3^q - 8 l_2^q - 14 l_1^q - (49/12)
\nonumber\\
a_{22}^M =\,& 17/8
\nonumber\\
a_{30}^M =\,& r_{M8}^q - 3l_{3}^{q}c_{9}^{q} - 64l_{3}^{q}c_{8}^{q}
  - 32l_{3}^{q}c_{7}^{q} - 128l_{3}^{q}c_{6}^{q}  + 32 c_{18}^{q}
   + 16c_{17}^{q} - 4c_{13}^{q} +24 c_{12}^{q} + 48 c_{11}^{q}
\nonumber\\&
 + 24c_{10}^{q} - 8c_{9}^{q} - 16c_{8}^{q} - 8c_{7}^{q} - 40 c_{6}^{q}
 + 12c_{5}^{q} + 4c_{4}^{q} - 8c_{3}^{q} - 7\left(l_{3}^{q}\right)^2
  - 22 l_{1}^{q}l_{3}^{q} - 4 l_{2}^{q}l_{3}^{q} 
\nonumber \\ &
 +\frac{157}{48} l_{3}^{q}+  \frac{8651}{1200} l_{2}^{q}
 + \frac{3823}{1200}l_{1}^{q} + \frac{4869659}{777600}
 - \frac{13}{6} \zeta _{3}
\nonumber \\
a_{31}^M =\,& 
 - 416 c_{18}^{q} - 208 c_{17}^{q} - 32 c_{16}^{q} + 96 c_{14}^{q}
  + 8 c_{13}^{q} - 48 c_{12}^{q} - 384 c_{11}^{q}
  - 192c_{10}^{q} + 72c_{9}^{q}  +  144 c_{8}^{q}
\nonumber\\&
  + 72 c_{7}^{q} + 64 c_{6}^{q} - 8 c_{5}^{q} - 56c_{4}^{q} + 16c_{3}^{q}
  + 32c_{2}^{q} - 96c_{1}^{q} - 8\left( l_{3}^{q}\right)^2
  - 48l_{2}^{q}l_{3}^{q} - 84l_{1}^{q}l_{3}^{q}  - \frac{88}{3} l_{3}^{q}
\nonumber\\&
 - \frac{231}{10}l_{2}^{q} - \frac{69}{5}l_{1}^{q} - \frac{74971}{8640}
\nonumber\\
a_{32}^M =\,&  \frac{23}{2} l_{3}^{q} - 11 l_{2}^{q} - 38 l_{1}^{q}
  - \frac{91}{24}
\nonumber\\
a_{33}^M =\,& \frac{103}{24}
\end{align}
where $\zeta _{k}$ is the Riemann-Zeta function. The leading log coefficient $a_{33}^M$ agrees with \cite{Bijnens:2009zi,Bijnens:2010xg}. 

The LECs $l_i^r$ are well known but the $c_i^r$ less well. In
\cite{Bijnens:1997vq,Bijnens:1999hw} combinations of the $p^6$ LECs appearing
at $p^6$ in $\pi\pi$-scattering, the mass and decay constant were defined,
$r_1,\ldots,r_6,r_M,r_F$, and numerical estimates using resonance saturation
were done
in \cite{Bijnens:1997vq}. The expressions in terms of the $c_i^r$ are given in
App.~\ref{Appri}.
We can check whether the $c_i^r$ dependence can be rewritten in terms of those.
This can be done for $a_{20}^M$ by definition and also for $a_{31}^M$.
However not completely for $a_{30}^M$ which in any case contains the free
$p^8$ LEC combination $r_{M8}^r$. Defining $r^q_i= 16\pi^2 r^r_i$ for $i=1,\ldots,6,M,F$, we obtain
\begin{align}
a_{30}^M =\,& r_{M8}^q - 4l_{3}^{q}r_F^q - 128l_{3}^{q}c_{6}^{q} 
  - r_6^q + 3 r_5^q + \frac{7}{2} r_4^q + \frac{1}{2} r_3^q + \frac{1}{2}r_M^q
  - 7\left(l_{3}^{q}\right)^2
  - 22 l_{1}^{q}l_{3}^{q} - 4 l_{2}^{q}l_{3}^{q} 
\nonumber \\ &
 +\frac{157}{48} l_{3}^{q}+  \frac{8651}{1200} l_{2}^{q}
 + \frac{3823}{1200}l_{1}^{q} + \frac{4869659}{777600}
 - \frac{13}{6} \zeta _{3}
\nonumber \\
a_{31}^M =\,& 
   - 6 r_6^q - 14 r_5^q - 11 r_4^q - 5 r_3^q - 2 r_2^q - \frac{5}{2} r_1^q
    + r_M^q - r_F^q
 - 8\left( l_{3}^{q}\right)^2
  - 48l_{2}^{q}l_{3}^{q} - 84l_{1}^{q}l_{3}^{q}  - \frac{88}{3} l_{3}^{q}
\nonumber\\&
 - \frac{231}{10}l_{2}^{q} - \frac{69}{5}l_{1}^{q} - \frac{74971}{8640}
\end{align}

\subsection{Decay constant}

For the decay constant everything is analogous except
that we need to evaluate the diagrams with one leg replaced by the axial
current and take into account the wave function renormalization factor $Z$.
Denoting the sum of one-particle-irreducible diagrams of the axial current as
$A(p^2=M_\pi^2)= A_{4}(p^2)+A_{6}(p^2)+A_{8}(p^2)$ the expression for the decay constant is (normalized to $1$ at lowest order)
\begin{align}
F_\pi =\,& F\sqrt{Z(M_\pi^2)} A(M_\pi^2)
\end{align}
Putting in the expanded expressions for $\Sigma$ and $A$ and using
$\partial^2\Sigma_4/(\partial p^2)^2=\partial A_4/\partial p^2=0$, we obtain
\begin{align}
F_4 =\,& \frac{1}{2}\frac{\partial\Sigma_4}{\partial p^2}+A_4
\nonumber\\
F_6 =\,& \frac{1}{2}\frac{\partial\Sigma_6}{\partial p^2}+
          +\frac{3}{8}\left(\frac{\partial\Sigma_4}{\partial p^2}\right)^2
          +\frac{1}{2}\frac{\partial\Sigma_4}{\partial p^2}A_4
          +A_6
\nonumber\\
F_8 =\,& A_{8} 
 +\Sigma _{4}\frac{\partial A_{6}}{\partial p^{2}}
 +\frac{5}{16}\left( \frac{\partial \Sigma _{4}}{\partial p^{2}}\right)^{3}
 +\frac{3}{8}\left[ A_{4}\left( \frac{\partial \Sigma _{4}}{\partial p^{2}}\right)^{2} +2\frac{\partial \Sigma _{4}}{\partial p^{2}}\frac{\partial \Sigma _{6}}{\partial p^{2}}\right]
\nonumber \\ &
 +\frac{1}{2}\left[ \frac{\partial \Sigma _{4}}{\partial p^{2}}A_{6} + \frac{\partial \Sigma _{6}}{\partial p^{2}}A_{4} + \Sigma_{4}\frac{\partial ^{2} \Sigma _{6}}{\left(\partial p^{2}\right)^{2}} +\frac{\partial \Sigma _{8}}{\partial p^{2}}\right]
\end{align}
with all right hand sides evaluated at $p^2=M^2$.

In order to obtain a finite result we need the subtraction
\begin{align}
r_{F8}^r =\,& (c\mu)^{-6\varepsilon}\Big\{ r_{F8}^r
        +\frac{\pi_{16}^3}{\varepsilon^3} (185/72)
        +\frac{\pi_{16}^3}{\varepsilon^2} (2117/432)
        -\frac{\pi_{16}^3}{\varepsilon} ( 20183/12960)
\nonumber\\&
        -\frac{\pi_{16}^2}{\varepsilon^2} \left( (659/72) l_{4}^r - 2 l_{3}^r - (1/6) l_{2}^r - (34/3) l_{1}^r\right)
\nonumber\\&
        -\frac{\pi_{16}^2}{\varepsilon} \left(  - (13/108) l_{4}^r - (53/12) l_{3}^r + (27/40) l_{2}^r + 
                  (61/15) l_{1}^r\right)
\nonumber\\&
        -\frac{\pi_{16}}{\varepsilon} ( 16 c_{20}^r + 64 c_{18}^r + 32 c_{17}^r - 8 c_{16}^r + 24 c_{14}^r
           + 2 c_{13}^r - 12 c_{12}^r + 96 c_{11}^r + 48 c_{10}^r + 8 c_{9}^r
\nonumber\\&
 + 16 c_{8}^r + 8 
           c_{7}^r - 48 c_{6}^r - 4 c_{5}^r - 28 c_{4}^r + 4 c_{3}^r + 8 c_{2}^r - 24 c_{1}^r - (7/2) 
           {l_{4}^r}^2 + 3 l_{3}^r l_{4}^r - (28/3) l_{2}^r l_{4}^r
\nonumber\\& - 16 l_{2}^r l_{3}^r - (44/3) l_{1}^r l_{4}^r - 28 l_{1}^r
            l_{3}^r) \Big\}
\end{align}

The results can be written in the form
\begin{align}
F_4 =\,& x\left(a_{10}^F+a_{11}^F L_M\right)
\nonumber\\
F_6 =\,& x^2\left(a_{20}^F+a_{21}^F L_M+a_{22}^F L_M^2\right)
\nonumber\\
F_8 =\,& x^3\left(a_{30}^F+a_{31}^F L_M+a_{32}^F L_M^2+a_{33}^F L_M^3\right)
\end{align}

The full results for the coefficients are, with $r^q_{F8}=(16\pi^2)^3r_{F8}^r$,
\begin{align}
a_{10}^F =\,& l_4^q
\nonumber\\
a_{11}^F =\,& -1
\nonumber\\
a_{20}^F =\,&  8 c_9^q + 16 c_8^q + 8 c_7^q
 - 2 l_3^q - l_2^q - (1/2) l_1^q - (13/192)
\nonumber\\
a_{21}^F =\,&  - (1/2) l_4^q - 2 l_3^q + 4 l_2^q + 7 l_1^q+ (23/12)
\nonumber\\
a_{22}^F =\,& -(5/4)
\nonumber\\
a_{30}^F =\,&  r_{F8}^{q} - 8 l_{4}^{q} c_{9}^{q} - 16 l_{4}^{q} c_{8}^{q}
 - 8 l_{4}^{q} c_{7}^{q} - 32 l_{4}^{q} c_{6}^{q}  - 64  c_{18}^{q}
 - 32  c_{17}^{q} +  c_{13}^{q} - 6  c_{12}^{q} 
          - 96  c_{11}^{q} - 48  c_{10}^{q}
\nonumber\\&
 + 16  c_{9}^{q} + 32 c_{8}^{q}  + 16 c_{7}^{q} + 40 c_{6}^{q} - 6  c_{5}^{q}
 - 2  c_{4}^{q} + 2  c_{3}^{q} - l_{3}^{q} l_{4}^{q}- 2  \left( l_{3}^{q}\right)^2
 +  l_{2}^{q} l_{4}^{q} + 4 l_{2}^{q} l_{3}^{q} 
+ \frac{1}{2} l_{1}^{q} l_{4}^{q}
\nonumber \\ &
 + 12 l_{1}^{q} l_{3}^{q} + \frac{313}{192} l_{4}^{q} + \frac{241}{48}  l_{3}^{q}
 + \frac{1469}{800} l_{2}^{q} + \frac{2359}{600} l_{1}^{q} - \frac{383293667}{1555200} + \frac{8}{9}  \zeta_{3}
\nonumber \\
a_{31}^F =\,& 
 - 16 c_{20}^{q} - 64 c_{18}^{q} - 32 c_{17}^{q} + 8 c_{16}^{q} - 24 c_{14}^{q}
 - 2 c_{13}^{q} + 12 c_{12}^{q} - 96 c_{11}^{q} - 48 c_{10}^{q} + 80 c_{6}^{q}
  + 4 c_{5}^{q}
\nonumber\\& 
          + 28 c_{4}^{q} - 4 c_{3}^{q}- 8 c_{2}^{q} + 24 c_{1}^{q}
   - l_{3}^{q} l_{4}^{q} - 4 l_{2}^{q} l_{4}^{q} + 16 l_{2}^{q} l_{3}^{q}
   - 7 l_{1}^{q} l_{4}^{q} + 28 l_{1}^{q} l_{3}^{q} - \frac{13}{6} l_{4}^{q}
   + \frac{17}{2} l_{3}^{q}
\nonumber\\&
  + \frac{569}{60} l_{2}^{q}
   + \frac{77}{10} l_{1}^{q} - \frac{7499}{2160}
\nonumber\\
a_{32}^F =\,&
  \frac{3}{8} l_{4}^{q} + \frac{27}{2} l_{2}^{q} + 33 l_{1}^{q}
  + \frac{1037}{144} 
\nonumber\\
a_{33}^F =\,&
   - \frac{83}{24}
\end{align}
Also here the leading log coefficient, $a_{33}^F$, agrees with
\cite{Bijnens:2010xg}.

We can similarly to the previous subsection rewrite the results
in terms of the combinations $r_i^q$ $i=1,\ldots,6,M,F$. The rewriting is
not fully possible here.
\begin{align}
a_{30}^F =\,&  r_{F8}^{q} - l_4^q r_F^q - 32 l_4^q c_6^q+ (1/2) r_6^q
  - (3/2) r_5^q - (3/2) r_4^q - (1/4) r_3^q - r_M^q + 4 c_{12}^q - 2 c_6^q
   - c_5^q 
\nonumber\\&- l_{3}^{q} l_{4}^{q}- 2  \left( l_{3}^{q}\right)^2
 +  l_{2}^{q} l_{4}^{q} + 4 l_{2}^{q} l_{3}^{q} 
+ \frac{1}{2} l_{1}^{q} l_{4}^{q}
 + 12 l_{1}^{q} l_{3}^{q} + \frac{313}{192} l_{4}^{q} + \frac{241}{48}  l_{3}^{q}
 + \frac{1469}{800} l_{2}^{q} + \frac{2359}{600} l_{1}^{q}
\nonumber\\&
 - \frac{383293667}{1555200} + \frac{8}{9}  \zeta_{3}
\nonumber \\
a_{31}^F =\,& 
   - (17/6) r_6^q + (33/2) r_5^q + (33/2) r_4^q + 3 r_3^q
          - (1/8) r_2^q - r_M^q - (3/2) r_F^q - 16 c_{20}^q - 20 c_{14}^q
\nonumber\\&
    - 96 c_{12}^q + 148 c_6^q - 46 c_5^q + (80/3) c_3^q
   - l_{3}^{q} l_{4}^{q} - 4 l_{2}^{q} l_{4}^{q} + 16 l_{2}^{q} l_{3}^{q}
   - 7 l_{1}^{q} l_{4}^{q} + 28 l_{1}^{q} l_{3}^{q} - \frac{13}{6} l_{4}^{q}
\nonumber\\&
   + \frac{17}{2} l_{3}^{q}
  + \frac{569}{60} l_{2}^{q}
   + \frac{77}{10} l_{1}^{q} - \frac{7499}{2160}
 \end{align}
 
\section{Numerical study:  mass dependence}

Now that the analytic forms of the mass and decay constant have been obtained,
we can do a first numerical analysis of the mass dependence. We present only
results for one choice of input parameters to give an impression of
the size of the NNNLO correction.

The expansions given in the previous section correspond to an expansion expressed in terms of the lowest order mass $M$ and decay constant $F$ in the form
\begin{align}
\label{xParamEqn}
M_{\pi}^{2} =\,& M^{2}\Big\{ 1+x\left(a_{10}^{M}+a_{11}^{M}L_{M}\right) +x^{2}\left(a_{20}^{M}+a_{21}^{M}L_{M}+a_{22}^{M}L_{M}^{2}\right)
\nonumber\\& +x^{3}\left(a_{30}^{M}+a_{31}^{M}L_{M}+a_{32}^{M}L_{M}^{2}+a_{33}^{M}L_{M}^{3}\right)
\Big\}
\nonumber\\
F_{\pi}=\,&F \Big\{ 1+x\left(a_{10}^{F}+a_{11}^{F}L_{M}\right) +x^{2}\left(a_{20}^{F}+a_{21}^{F}L_{M}+a_{22}^{F}L_{M}^{2}\right)
\nonumber\\& +x^{3}\left(a_{30}^{F}+a_{31}^{F}L_{M}+a_{32}^{F}L_{M}^{2}+a_{33}^{F}L_{M}^{3}\right)
\Big\}
\end{align}
with $x=M^{2}/(16\pi ^{2}F^{2})$ and $L_{M} = \log( M^{2}/\mu ^{2})$.

There are many ways to rewrite this expansion but the second most standard
version is the inverse, namely
\begin{align}
\label{xiParamEqn}
M^{2} =\,& M_{\pi}^{2}\Big\{ 1+\xi \left(b_{10}^{M}+b_{11}^{M}L_{\pi}\right) +\xi ^{2}\left(b_{20}^{M}+b_{21}^{M}L_{\pi}+b_{22}^{M}L_{\pi}^{2}\right)
\nonumber\\&
 +\xi ^{3}\left(b_{30}^{M}+b_{31}^{M}L_{\pi}+b_{32}^{M}L_{\pi}^{2}+b_{33}^{M}L_{\pi}^{3}\right)
\Big\}
\nonumber\\
F=\,&F_{\pi} \Big\{ 1+\xi \left(b_{10}^{F}+b_{11}^{F}L_{\pi }\right) +\xi ^{2}\left(b_{20}^{F}+b_{21}^{F}L_{\pi}+b_{22}^{F}L_{\pi}^{2}\right)
\nonumber\\& +\xi ^{3}\left(b_{30}^{F}+b_{31}^{F}L_{\pi}+b_{32}^{F}L_{\pi}^{2}+b_{33}^{F}L_{\pi}^{3}\right)
\Big\}
\end{align}
where $\xi=M_{\pi}^{2}/(16\pi ^{2}F_{\pi}^{2})$ and $L_{\pi} = \log( M_{\pi}^{2}/\mu ^{2})$. The analytic expressions of the $a_{ij}^{F.M}$ are in the previous
section and the $b_{ij}^{F,M}$ can be found in App.~\ref{AnalyticCoeffApp}.
These are often referred to as the $x$- and $\xi$-expansion,
see e.g. \cite{FLAG}.

As input we use $\mu = 0.77$~GeV, $\bar l_1 = -0.4, \bar l_2 = 4.3$
from \cite{Colangelo:2001df}, $\bar l_3=3.41,\bar l_4=4.51$ from the $N_f=2$
estimates of \cite{FLAG}. The numerical values for the $r_i^r$
are taken from \cite{Bijnens:1997vq}
\begin{align}
10^4 r_1^r =\,&-0.6 & 10^4 r_2^r =\,& 1.3 & 10^4 r_3^r =\,& -1.7
& 10^4 r_4^r =\,&-1.0 
\nonumber\\
 10^4 r_5^r =\,& 1.1 & 10^4 r_6^r =\,& 0.3 &
r_M^r =\,& 0 & r_F^r =\,&0
\end{align}
The remaining $c_i^r$ and $r_{M8}^r$, $r_{F8}^r$ have been set to zero.
The resulting numerical values of the $a_{ij}^{F.M}$ and $b_{ij}^{F,M}$ can be found in Table~\ref{NumParamsTable}.
\begin{table}
\caption{Numerical values of the $a_{ij}^{M,F}$ and $b_{ij}^{M,F}$ for the input
parameters given in the text.}
\begin{center}
 \begin{tabular}{|c||c |c |c|c|} 
 \hline
 $ij$ & $a_{ij}^{M}$ & $b_{ij}^{M}$ & $a_{ij}^{F}$ & $b_{ij}^{F}$ \\ [0.7ex]
 \hline
 \hline
 10 & 0.0028 & $-$0.0028 & 1.0944 & $-$1.0944\\ 
 \hline
 11 & 0.5 & $-$0.5 & $-$1.0 &1.0\\
 \hline
 20 & 1.6530 &  $-$1.6577& $-$0.0473 & $-$1.1500\\
 \hline
 21 & 2.4573 & $-$3.2904 & $-$1.9058 & 4.1388\\
 \hline
 22 & 2.125 & $-$0.625 & $-$1.25 & $-$0.25\\ 
 \hline
  30 & 0.4133 & $-$6.8035 & $-$244.5350 & 242.2724 \\
 \hline
   31 &$-$3.7044 &  4.2718&  $-$15.4989& 28.5703\\
  \hline
    32 & 17.1476 & 0.6204& $-$9.3946 & $-$6.7751\\
\hline
 33 & 4.2917 &  5.1458&  $-$3.4583 & $-$0.4167\\
 \hline
\end{tabular}
\end{center}\label{NumParamsTable}
\end{table}
Note that the numerical values of $a_{30}^F$ and $b_{30}^F$ are rather large.
This is due to the very large numerical
coefficient $383293667/1555200\approx 246.5$ appearing there, the remaining
coefficients are of more natural size.
 
\begin{figure}[htbp]
\begin{minipage}{0.48\textwidth}
\includegraphics[width=0.99\textwidth]{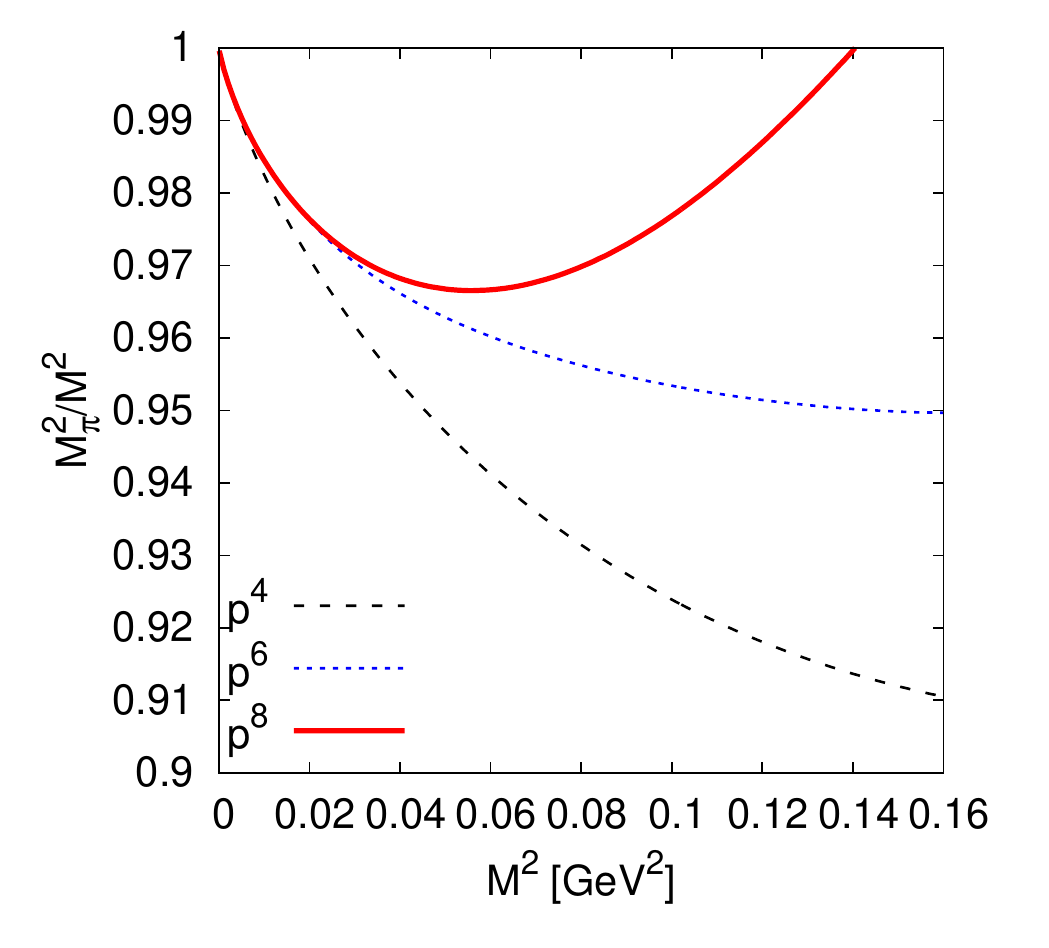}\\[-3mm]
\centerline{(a)}
\includegraphics[width=0.99\textwidth]{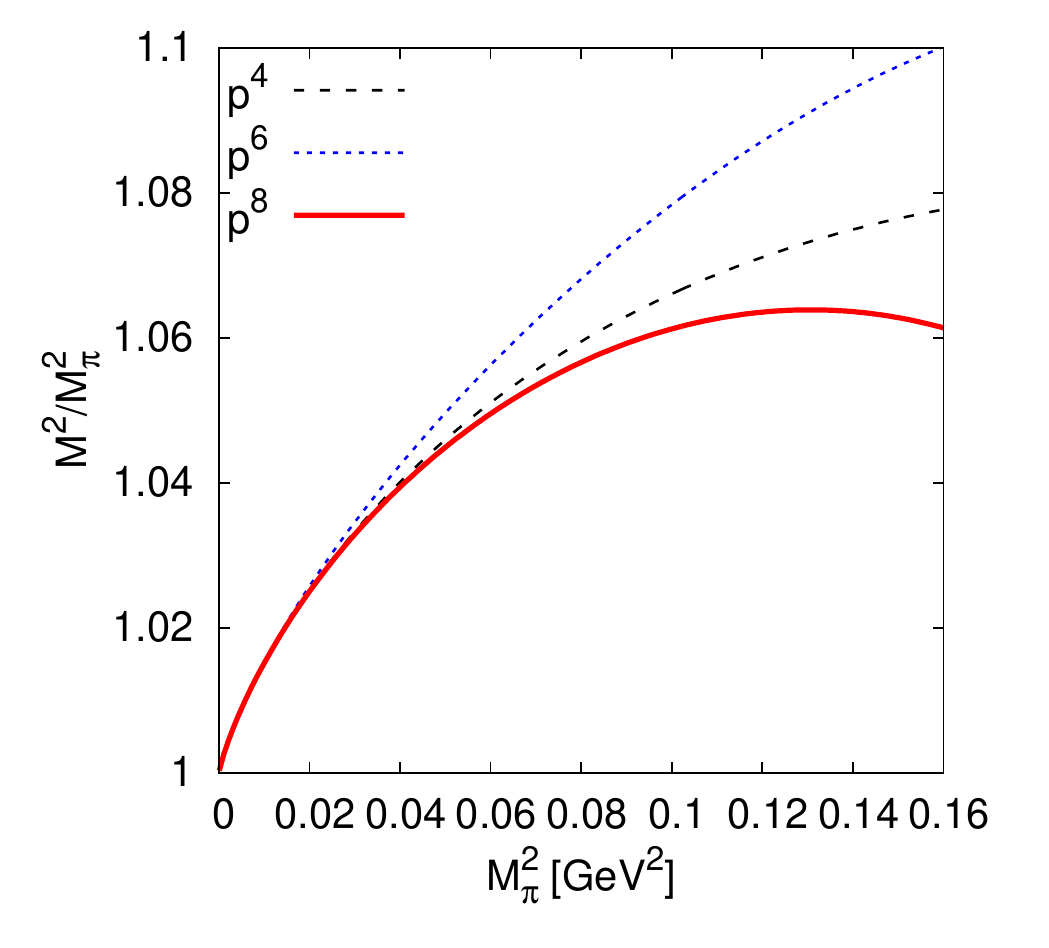}\\[-3mm]
\centerline{(c)}
\end{minipage}
\begin{minipage}{0.48\textwidth}
\includegraphics[width=0.99\textwidth]{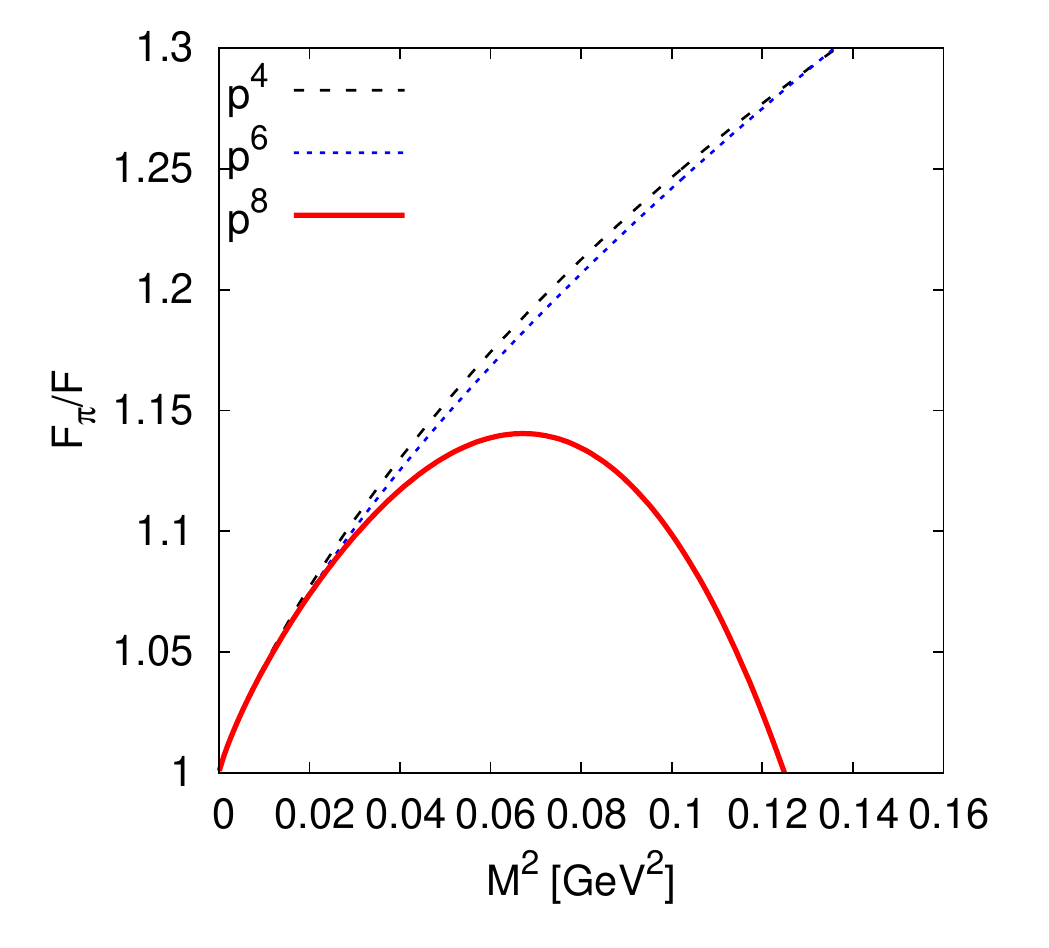}\\[-3mm]
\centerline{(b)}
\includegraphics[width=0.99\textwidth]{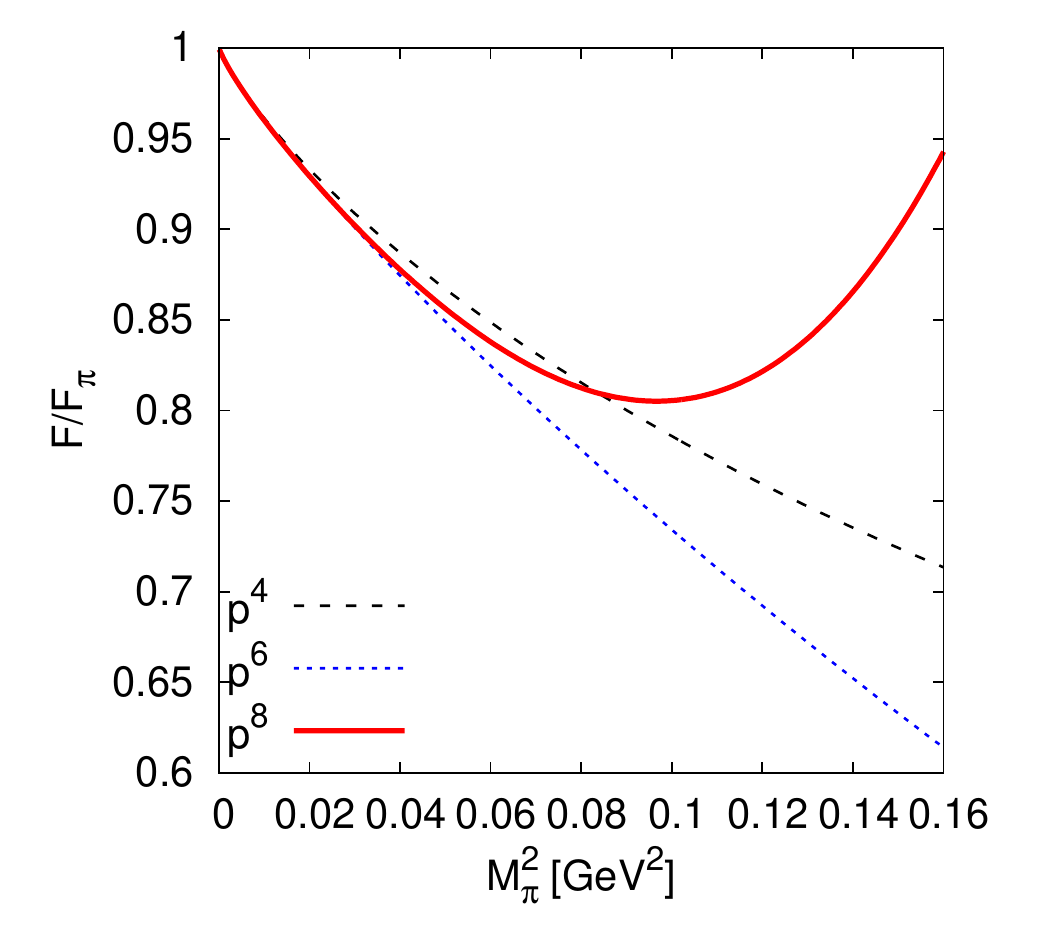}\\[-3mm]
\centerline{(d)}
\end{minipage}
\caption{(a)~The $x$-expansion for the mass, (b)~the $x$-expansion for the
decay constant, (c)~the $\xi$-expansion for the mass, (d)~the $\xi$-expansion
for the decay constant at NLO, NNLO and NNNLO. LO is constant at 1 for all four
plots.}
\label{figplots}
\end{figure}

The quantities in (\ref{xParamEqn})--(\ref{xiParamEqn}) are plotted in 
Fig.~\ref{figplots}(a--d), with the same inputs as above.
For the $\xi$-expansion we kept $F_\pi= 92.2$~MeV constant while varying $M_\pi$
and for the $x$-expansion we kept $F= 92.2/1.073$~MeV constant while
varying $M$. The convergence around the physical value
$M_\pi^2\approx0.02$~GeV$^2$ is excellent.
For the mass, the $\xi$-expansion converges much better, for the decay constant
it is somewhat better. The effect of the very large constants
$a_{30}^F$ and $b_{30}^F$ is clearly visible in the results for the decay
constant.

\section{Conclusions}

In this paper we presented the calculation of the NNNLO contributions to
the pion mass and decay constant in the isospin limit of two-flavour ChPT.
We also calculated the mass in the $O(N)$ $\phi^4$ case to show the principle.
This required the evaluation of $2\times 34$ diagrams and their derivatives
w.r.t.\ the momentum. The master integrals needed for the calculation were
known. The tree level contributions from the NNNLO Lagrangian are unknown, but
were here parameterized as free renormalized parameters. We reproduced the
known NNLO results and the known leading logarithms.

A small numerical study of the two quantities was performed, and there was
continuing good convergence at the physical pion mass.

\section*{Acknowledgements}

This work is supported in part by the Swedish Research Council grants
contract numbers 621-2013-4287, 2015-04089 and 2016-05996 and by
the European Research Council (ERC) under the European Union's Horizon 2020
research and innovation programme (grant agreement No 668679).

\appendix
\section{Master integrals}
\label{IntApp}

Below, the master integrals needed for the calculation\footnote{In this section, everything has been truncated so as to only include terms relevant for this particular calculation.} are listed.
To simplify the expressions of the master integrals below, we define 
\begin{align}
C\left( \varepsilon ,\, M^{2}\right) =\,&
(4\pi)^\varepsilon \Gamma(1+\varepsilon) M^{-2\varepsilon}\,.
\end{align}
Also, we denote the external momentum as $p$
with $p^2=M^2$, as is relevant for the quantities considered here.

\subsection{One loop}

We only need one one-loop integral, the well-known tadpole:
\begin{align}
\frac{1}{i}\int \frac{d^d k}{(2\pi)^d}\frac{1}{k^2-M^2}
=\,& \frac{C\left( \varepsilon , \, M^{2}\right)}{16\pi ^{2}} M^{2}\left( \frac{1}{\varepsilon} + 1+\varepsilon + \varepsilon ^{2}\right)  
\end{align}

\subsection{Two loops}

All needed two-loop integrals can be reduced to the equal mass on-shell sunset
or products of tadpoles. The sunset result is, see e.g. \cite{Laporta:2004rb}
\begin{align}
\frac{1}{i^2}\int \frac{d^d k}{(2\pi)^d}\frac{d^d l}{(2\pi)^d}
\frac{1}{(k^2-M^2)(l^2-M^2)((p-k-l)^2-M^2)}
 =\,&
\nonumber\\&
\hskip-40ex
\left( \frac{C\left( \varepsilon , \, M^{2}\right) }{16 \pi ^{2}}\right)^{2}M^{2}
\left[ \frac{3}{2}\frac{1}{\varepsilon ^{2}}+\frac{17}{4}\frac{1}{\varepsilon} + \frac{59}{8}+\left( \frac{65}{16}+8\zeta _{2}\right) \varepsilon \right] 
\end{align}

\subsection{Three loops}

We only need two more master integrals at three-loop order. The other combinations
are products of the one- and two-loop integrals.
The first one is a vacuum integral:
\begin{align}
\frac{1}{i^3}\int \frac{d^d k}{(2\pi)^d}\frac{d^d l}{(2\pi)^d}\frac{d^d q}{(2\pi)^d}
\frac{1}{\left( k^{2}-M^{2}\right) \left( (k+l)^{2}-M^{2}\right)
 \left((l+q)^{2}-M^{2}\right) \left( q^{2}-M^{2}\right)} = &
\nonumber\\
\hskip-40ex
  \left( \frac{C\left( \varepsilon , \, M^{2}\right) }{16 \pi ^{2}}\right) ^{3}M^{4}\left[ 2\frac{1}{\varepsilon ^{3}}+\frac{23}{3}\frac{1}{\varepsilon ^{2}}+\frac{35}{2}\frac{1}{\varepsilon}+\frac{275}{12}\right]\,.
\end{align}
The second needed three-loop integral has external momentum running through it
\begin{align}
\frac{1}{i^3}\int \frac{d^d k}{(2\pi)^d}\frac{d^d l}{(2\pi)^d}\frac{d^d q}{(2\pi)^d}
\frac{1}{\left( k^{2}-M^{2}\right) \left( (k+l)^{2}-M^{2}\right)
  \left( (p-l)^{2}-M^{2}\right)\left((l+q)^{2}-M^{2}\right) \left( q^{2}-M^{2}\right)} = &
\nonumber\\&
\hskip-70ex
 \left( \frac{C\left( \varepsilon , \, M^{2}\right) }{16 \pi ^{2}}\right) ^{3}M^{2}\left[ \frac{1}{\varepsilon^{3}}+\frac{16}{3}\frac{1}{\varepsilon ^{2}}+16\frac{1}{\varepsilon}+20-2\zeta _{3}+16\, \zeta _{2}\right]
\end{align}
These two are denoted as $I_{12}$ and $I_{8}$, respectively, in~\cite{Melnikov:2000zc}.

\section{The expressions for the $r_i^r$}
\label{Appri}

These are reproduced from \cite{Bijnens:1999hw}.
\begin{align}
r_F^r =\,& 8 c_{7}^r+16 c_{8}^r+8 c_{9}^r
\nonumber\\
r_M^r =\,& -32 c_{6}^r-16 c_{7}^r-32 c_{8}^r-16 c_{9}^r+48 c_{10}^r+96 c_{11}^r+32 
c_{17}^r+64 c_{18}^r
\nonumber\\
r_1^r =\,& 64 c_{1}^r-64 c_{2}^r+32 c_{4}^r-32 c_{5}^r+32 c_{6}^r-64 c_{7}^r-128 c_{8}^r-64 c_{9}^r
 +96 c_{10}^r+192 c_{11}^r-64 c_{14}^r
\nonumber\\&
+64 c_{16}^r+96 c_{17}^r+192 c_{18}^r
\nonumber\\
r_2^r =\,& -96 c_{1}^r+96 c_{2}^r+32 c_{3}^r-32 c_{4}^r+32 c_{5}^r-64 c_{6}^r+32 c_{7}^r+64 c_{8}^r+32 c_{9}^r
 -32 c_{13}^r+32 c_{14}^r
\nonumber\\&
-64 c_{16}^r
\nonumber\\
r_3^r =\,& 48 c_{1}^r-48 c_{2}^r-40 c_{3}^r+8 c_{4}^r-4 c_{5}^r+8 c_{6}^r-8 c_{12}^r+20 c_{13}^r
\nonumber\\
r_4^r =\,& -8 c_{3}^r+4 c_{5}^r-8 c_{6}^r+8 c_{12}^r-4 c_{13}^r
\nonumber\\
r_5^r =\,& -8 c_{1}^r+10 c_{2}^r+14 c_{3}^r
\nonumber\\
r_6^r =\,& 6 c_{2}^r+2 c_{3}^r
\end{align}

\section{Analytic expressions for $b_{ij}^{M,F}$ }
\label{AnalyticCoeffApp}

Below, the analytic forms of the coefficients $b_{ij}^{M,F}$ are given.
The notation is the same as in Sect.~\ref{sec:analytical}.
We only give here the result in terms of the $r_i^r$.

The mass coefficients are
\begin{align}
b_{10}^{M} =\,& -2l_{3}^{r}
\nonumber\\
b_{11}^{M} =\,& -\frac{1}{2}
\nonumber\\
b_{20}^{M} =\,& - \frac{163}{96} - r^{q}_{M} - 4 l_{3}^{q} l_{4}^{q} + 8 \left( l_{3}^{q}\right) ^{2} - 2 l_{2}^{q} - l_{1}^{q}
\nonumber\\
b_{21}^{M} =\,& \frac{13}{3} - l_{4}^{q} + 11 l_{3}^{q} + 8 l_{2}^{q} + 14 l_{1}^{q}
\nonumber\\
b_{22}^{M} =\,& -\frac{5}{8}
\nonumber\\
b _{30}^{M} =\,&  - \frac{4209509}{777600} + r_{q}^{6} - 3 r_{q}^{5} - \frac{7}{2} r_{q}^{4} - \frac{1}{2} r_{q}^{3} + \frac{13}{6}
       \zeta _{3} - \frac{163}{24} l_{4}^{q} - 4 l_{4}^{q} r_{q}^{M} + \frac{93}{16} l_{3}^{q} + 10 l_{3}^{q} r_{q}^{M} 
\nonumber \\ &
 + 128 
      l_{3}^{q} c_{6}^{q} - 10 l_{3}^{q} \left( l_{4}^{q}\right) ^{2} + 40 \left( l_{3}^{q}\right) ^{2} l_{4}^{q} - 40 \left( l_{3}^{q}\right) ^{3}
       - \frac{7451}{1200} l_{2}^{q} - 8 l_{2}^{q} l_{4}^{q} + 12 l_{2}^{q} l_{3}^{q} - \frac{3223}{1200} l_{1}^{q}
\nonumber \\ &
  - 4
       l_{1}^{q} l_{4}^{q} + 6 l_{1}^{q} l_{3}^{q}
\nonumber\\
b _{31}^{M} =\,&   \frac{134551}{8640} + 6 r_{q}^{6} + 14 r_{q}^{5} + 11 r_{q}^{4} + 5 r_{q}^{3} + 2 r_{q}^{2} + \frac{5}{2} 
      r_{q}^{1} + \frac{11}{2} r_{q}^{M} + \frac{52}{3} l_{4}^{q} - \frac{5}{2} \left( l_{4}^{q}\right) ^{2} 
 - \frac{56}{3} l_{3}^{q}
\nonumber \\ &
 + 54 l_{3}^{q} l_{4}^{q} - 84 \left( l_{3}^{q}\right) ^{2} + \frac{291}{10} l_{2}^{q} + 32 l_{2}^{q} l_{4}^{q} - 48 l_{2}^{q} 
      l_{3}^{q} + \frac{34}{5} l_{1}^{q} + 56 l_{1}^{q} l_{4}^{q} - 84 l_{1}^{q} l_{3}^{q}
\nonumber\\
b _{32}^{M} =\,& - \frac{71}{3}  - \frac{1}{2}  l_{4}^{q} - \frac{161}{4}  l_{3}^{q} - 45 l_{2}^{q} - 60 l_{1}^{q}
\nonumber\\
b _{33}^{M} =\,& \frac{247}{48} 
\end{align}

The coefficients for the decay constant are
\begin{align}
b_{10}^{F} =\,& -l_{4}^{q}
\nonumber\\
b_{11}^{F} =\,& 1
\nonumber\\
b_{20}^{F} =\,&  \frac{13}{192} - r_{q}^{F} - \left( l_{4}^{q}\right) ^{2} + 2 l_{3}^{q} l_{4}^{q} + l_{2}^{q} + \frac{1}{2} l_{1}^{q}
\nonumber\\
b_{21}^{F} =\,& - \frac{29}{12} + 3 l_{4}^{q} - 4 l_{2}^{q} - 7 l_{1}^{q}
\nonumber\\
b_{22}^{F} =\,& -\frac{1}{4} 
\nonumber\\
b _{30}^{F} =\,& \frac{380653067}{1555200} - \frac{1}{2} r_{q}^{6} + \frac{3}{2} r_{q}^{5} + \frac{3}{2} r_{q}^{4} + \frac{1}{4} 
      r_{q}^{3} - \frac{8}{9} \zeta _{3} - 4 c_{12}^{q} + 2 c_{6}^{q} + c_{5}^{q} + \frac{65}{192} l_{4}^{q} + l_{4}^{q} r_{q}^{M}
\nonumber \\ & 
       - 3 l_{4}^{q} r_{q}^{F} + 32 l_{4}^{q} c_{6}^{q} - 2 \left(l_{4}^{q} \right) ^{3} - \frac{35}{24} l_{3}^{q} + 4 l_{3}^{q} 
      r_{q}^{F} + 8 l_{3}^{q} \left( l_{4}^{q}\right) ^{2} - 8 \left( l_{3}^{q}\right) ^{2} l_{4}^{q} - \frac{3069}{800} l_{2}^{q} + 5 
      l_{2}^{q} l_{4}^{q}
\nonumber \\ &
  - \frac{2959}{600} l_{1}^{q} + \frac{5}{2} l_{1}^{q} l_{4}^{q}
\nonumber\\
b _{31}^{F} =\,& \frac{58121}{8640} + \frac{17}{6} r_{q}^{6} - \frac{33}{2} r_{q}^{5} - \frac{33}{2} r_{q}^{4} - 3 r_{q}^{3} + \frac{1}{8} r_{q}^{2} + \frac{13}{2} r_{q}^{F} + 16 c_{20}^{q} + 20 c_{14}^{q} + 96 c_{12}^{q} 
\nonumber \\ &
 - 148 c_{6}^{q}
       + 46 c_{5}^{q} - \frac{80}{3} c_{3}^{q} - \frac{145}{12} l_{4}^{q} + 10 \left( l_{4}^{q}\right) ^{2} - \frac{5}{6} 
      l_{3}^{q} - 16 l_{3}^{q} l_{4}^{q} - \frac{389}{60} l_{2}^{q} - 20 l_{2}^{q} l_{4}^{q}
 \nonumber \\ &
 + \frac{63}{10} l_{1}^{q}
 - 35 l_{1}^{q} l_{4}^{q}
\nonumber\\
b _{32}^{F} =\,&  \frac{859}{144}  - \frac{25}{4}  l_{4}^{q} + \frac{29}{2}  l_{2}^{q} + 16 l_{1}^{q}
\nonumber\\
b _{33}^{F} =\,& - \frac{5}{12} 
\end{align}

\bibliography{pionmassdecay}
\bibliographystyle{JHEPmy}

\end{document}